\documentclass[journal]{IEEEtran}

\usepackage{slashbox}
\usepackage{lscape}
\usepackage{multirow}
\usepackage{graphicx}
\usepackage{caption}
\usepackage{subcaption}
\usepackage{xcolor}
\usepackage{amssymb}
\usepackage{newtxtext,newtxmath}
\usepackage{algorithm}
\usepackage[noend]{algpseudocode}
\usepackage{tabularx,ragged2e,booktabs}
\usepackage{tabulary}
\usepackage{bm}
\usepackage{cite}
\usepackage{soul}
\usepackage{hyperref}

\usepackage{amsmath}

\newcolumntype{L}{>{\RaggedRight\arraybackslash}X}

\ifCLASSINFOpdf
 
\else
  
\fi

\begin{document}

\title{A Directional Vibrotactile Feedback Interface for Ergonomic Postural Adjustment}

\author{Wansoo Kim$^{+*}$, \IEEEmembership{Member, IEEE}, Virginia Ruiz Garate$^+$, \IEEEmembership{Member, IEEE}, Juan M. Gandarias, \IEEEmembership{Member, IEEE}, \\Marta Lorenzini, \IEEEmembership{Member, IEEE}, and Arash Ajoudani \IEEEmembership{Member, IEEE}
\thanks{The authors are with the HRI$^{2}$ Lab, Istituto Italiano di Tecnologia, Genoa, Italy.}
\thanks{This work was supported in part by the ERC-StG Ergo-Lean (Grant Agreement No.850932), in part by the European Union’s Horizon 2020 research and innovation programme under Grant Agreement No. 871237 (SOPHIA).}
\thanks{W. Kim is also with Hanyang University, South Korea (e-mail: wansookim@hanyang.ac.kr). }
\thanks{V. Ruiz is also with University of the West of England, United Kingdom (e-mail: Virginia.RuizGarate@uwe.ac.uk). }
\thanks{*Corresponding author.}
\thanks{$^{+}$ Contributed equally to this work.}}

\markboth{Now published in IEEE Transactions on Haptics DOI:  \href{http://doi.org/10.1109/TOH.2021.3112795}{10.1109/TOH.2021.3112795}} %
{KIM \MakeLowercase{\textit{et al.}}: A Directional Vibrotactile Feedback Interface for Ergonomic Postural Adjustment}

\maketitle

\begin{abstract}
The objective of this paper is to develop and evaluate a directional vibrotactile feedback interface as a guidance tool for postural adjustments during work. In contrast to the existing active and wearable systems such as exoskeletons, we aim to create a lightweight and intuitive interface, capable of guiding its wearers towards more ergonomic and healthy working conditions. 
To achieve this, a vibrotactile device called ErgoTac is employed to develop three different feedback modalities that are able to provide a directional guidance at the body segments towards a desired pose. In addition, an evaluation is made to find the most suitable, comfortable, and intuitive feedback modality for the user. Therefore, 
these modalities are first compared experimentally on fifteen subjects wearing eight ErgoTac devices to achieve targeted arm and torso configurations. The most effective directional feedback modality is then evaluated on five subjects in a set of experiments in which an ergonomic optimisation module provides the optimised body posture while performing heavy lifting or forceful exertion tasks. The results yield strong evidence on the usefulness and the intuitiveness of one of the developed modalities in providing guidance towards ergonomic working conditions, by minimising the effect of an external load on body joints. We believe that the integration of such low-cost devices in workplaces can help address the well-known and complex problem of work-related musculoskeletal disorders.
\end{abstract}


\IEEEpeerreviewmaketitle

\section{Introduction}

\IEEEPARstart{M}{usculoskeletal} disorders (MSDs) in workplaces are the leading cause of injuries and employee absenteeism, threatening workers' well-being and causing dramatic losses of productivity in industrial countries. In European Union, the total associated cost is estimated to be around 2\% of gross domestic product (GDP) \cite{bevan2015economic}.

Many attempts have been made to tackle this complex issue from different angles, from the ergonomic design of the workstations \cite{ajoudani2018progress, bossomaier2010scientific, ben2002ergonomic}, to the development of assistive \cite{lim2015development, peternel2014teaching, amor2014interaction, evrard2009teaching}, and warning/feedback devices \cite{yan2017wearable, vignais2013innovative}. While the first two have received considerable attention, the last, i.e., the development of intuitive feedback systems that can warn workers about inappropriate working postures, is often disregarded or considered as a secondary concern. This is however in contrast to the underlying cost-profit trade-off, since such feedback devices can be low-cost and very effective in minimising risks to workers' health. 

\begin{figure}
    \centering
    \includegraphics[width=0.9 \columnwidth]{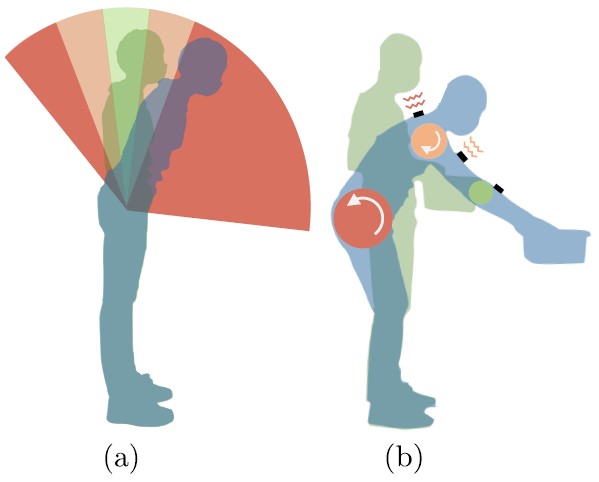}
    \caption{The proposed haptic interface provides different vibration levels and directional feedback to achieve an optimal configuration in which the effect of external loads on body joints is minimum. (a) The vibration level conveys the distance from the current (blue) to the target (green) configurations. Here, the colours define the vibration level: red--high, orange--medium, green--no vibration. (b) The vibrotactile devices are attached to multiple joints to provide haptic guidance considering the correct direction to achieve the desired configuration.}
    \label{fig:my_label}
\end{figure}

Despite this substantial advantage, industrial working environments pose a number of challenges to the choice and use of such feedback modalities. First, the most straightforward visual feedback \cite{kim2019adaptable,zhang2006evaluation} is not advisable, due to the amount of distraction caused by staring at the screens. The use of audio feedback systems is also not recommended \cite{goomas2010ergonomics} because of the high levels of occupational noise, i.e., the high amount of acoustic energy received by an employee's auditory system, typical in industrial environments. Mechano-tactile feedback devices \cite{ajoudani2014exploring, fan2008haptic} as well may cause discomfort in prolonged industrial operations due to the applied mechanical pressure, and could result in low acceptability among their users.

The second challenge concerns the possibility to provide feedback to multiple body segments \cite{dunkelberger2018improving}, since several industrial tasks involve articulated  movements. This requirement eliminates the choice of bulky and heavy wearable systems that become unusable when several joints/links are involved. The third challenge is associated with the learning and familiarisation time of the feedback devices, which should be minimised. Excluding the aforementioned  modalities unsuitable for industrial environments, and taking into account the wearability and familiarisation constraints, vibrotactile feedback appears to be a promising choice \cite{alahakone2009vibrotactile}. In fact, the application of vibrotactile displays for operators' awareness in human-robot interaction \cite{casalino2018operator}, balance control \cite{ballardini2020vibrotactile}, prosthetic control \cite{ajoudani2014exploring}, and 
teleoperation \cite{debus2002multi} have already shown promising results. Due to their low-cost and small size, vibrotactile displays have also found their way to multi-point feedback, e.g., for posture optimisation and teleoperation control \cite{zheng2010vibrotactile,debus2002multi}.

From the application point of view, several approaches provide the segments' directional information through haptic guidance and evaluate their impact. For example, enabling the perception of directional cues through the use of a single actuator \cite{chen2016lower}, comparing feedback accuracy via lower-fidelity (4-motor) and higher-fidelity (8-motor) on the wrist-based vibro-motor feedback \cite{hong2016evaluating}, employing the asymmetrically accelerated mass to provide high-resolution directional haptic cues in multiple dimensions \cite{tappeiner2009good}, and multi-sensory haptic cue (i.e., interface of stretch, squeeze, and integrated vibration feedback modalities) \cite{dunkelberger2018improving} are some approaches that have succeeded to provide the directional haptic information to the user (or wearer). 

This paper presents the development and experimental evaluation of a real-time haptic feedback interface with directional guidance for optimal ergonomic posture adjustment (see Fig.1). The core of the interface is the low-cost, wearable, vibrotactile device ErgoTac, presented in our previous work \cite{kim2018ergotac}. The main and significant differences of this research study w.r.t. our previous work are the implementation of the directional vibrotactile guidance and the evaluation of different vibrotactile feedback modalities by the multiple subjects. The former is essential to give the worker useful feedback about the desired movement. Indeed, the use of multiple ErgoTac devices on different body parts allows to give haptic feedback to the wearer based on three statements: i) which joints are not in the desired configuration, ii) how far a particular joint is from the target position, iii) in which direction that joint has to move to approach the desired position. Furthermore, several vibrotactile feedback modalities are implemented into the ErgoTac devices, hence we are able to assess intuitiveness of each vibrotactile stimulus.

To the best of our knowledge, this is the first approach that provides and evaluates the performance of an intuitive, directional vibrotactile feedback interface for ergonomic guidance, considering not only the kinodynamic aspect but also the task constraints (i.e., requirements). Therefore, the contribution of this work is two-folded:

\begin{itemize}
    \item First, to find the most suitable, comfortable, and intuitive vibration modality for the user. In this regard, three modalities are proposed: PATTERN, SPOT, and RAMP.
    \item Second, to integrate the best feedback modality with an ergonomic optimisation framework. This way, directional haptic guidance can be given to a worker who has a non-ergonomic configuration.
\end{itemize}

For both contributions, an experimental evaluation has been carried out considering multiple subjects of different age, gender and complexion, and a statistical analysis has been performed evaluating objective and subjective metrics.

The  rest  of  the  paper  is  organised  as  follows. Section II describes the ErgoTac device, and the feedback methods. Section III details the Ergonomic framework. In Section IV, the experimental evaluation and results are described. The outcomes of this work are discussed in Section V. Finally, Section VI presents the conclusions and prospective research work.

\section{Materials and Methods}

\subsection{Ergotac: a tactile feedback interface}
\begin{figure}
    \centering
    \includegraphics[width = 0.9\columnwidth]{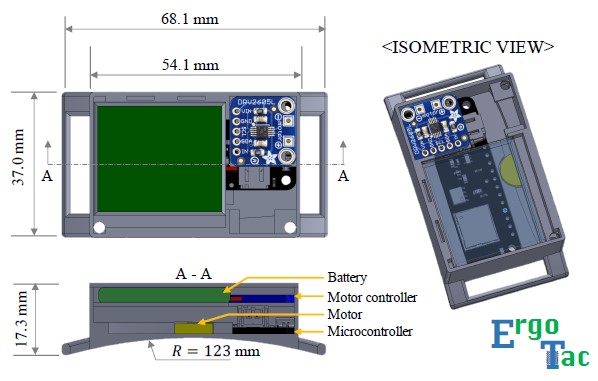}
    \caption{The specification and components of the developed wearable vibrotactile feedback device ErgoTac.}
    \label{fig:ergotac}
\end{figure}

The core component of the interface proposed in this paper is the new vibrotactile feedback device ErgoTac, which we presented in \cite{kim2018ergotac}. Fig. \ref{fig:ergotac} illustrates the CAD models of the ErgoTac device along with its components and specifications. 
ErgoTac is designed as a wireless vibrotactile device placed on the human body's segments to warn the user when exceeding ergonomic indicators during physically demanding tasks.
The dimension of the ErgoTac device is 68.1 mm $\times$ 37.0 mm $\times$ 17.3 mm, and the weight is 28 g. 
A mini-eccentric rotating mass (ERM) vibration motor with 10 mm diameter and 2.7 mm thickness is mounted on the bottom inside of the box. An ERM vibration motor can produce a varying frequency of vibration up to approximately 121 Hz, depending on the operating voltage. The frequency of the vibration on ErgoTac device is set to 121 Hz to avoid a tendon vibration illusion effect \cite{kammers2006dissociating}.
Moreover, the wireless communication protocol supports multi-point connection via Bluetooth low energy (2.4 GHz). 
Hence, the device size, weight, and communication make it ideal for applications in which wearability is required. 
ErgoTac can provide three different vibrotactile amplitudes determined by the level of danger from the ergonomics perspective. The wearer can feel the vibration strength differences in multiple ErgoTacs and take actions accordingly, so as to minimise or avoid non-ergonomic configurations. In \cite{kim2018ergotac}, the developed vibrotactile framework provides quantitative warnings regarding the overloading induced on the body joint by an external load, which is considered as the ergonomic indicator. However, users did not receive any feedback about the optimal postural adjustment (i.e., the framework did not provide directional guidance). 

In this research, we build on our previous approach and present several feedback methods that address directional guidance to the user. These different feedback modalities are developed to provide adequate haptic guidance, including directional information, and ensure swift and intuitive feedback toward an optimise the ergonomic configuration.

\subsection{Vibrotactile feedback modalities}
\label{subsec:feedback_methods}
To find the best approach to convey the desired postural information with the external devices (i.e., ErgoTac), the user's understanding of the provided feedback and the required cognitive effort should be studied. With this aim, we first consider three aspects of the vibrotactile feedback (i.e., intensity, duration and position) that influence the modalities to be used:
\begin{itemize}
    \item \textit{Intensity of the vibrations.} Vibration intensity was observed to have a significant effect on user experience. The risk of vibration-associated injuries increases with both the intensity and duration of the vibration exposure \cite{tan2019user}, however, higher intensities are easier to be detected \cite{Shull2015}. Since we are targeting work environments where the users will use the device for several hours, low intensity vibrations must be used to maximise sustainability.
    \item \textit{Duration of the vibrations.} Research has demonstrated that as short as 100 ms the vibrations can be perceived while avoiding overlap \cite{Filosa2019,Crea2017}. Nonetheless, longer duration is needed when lower intensity is used \cite{Li2015}. ErgoTac vibrations originally had a duration of 200 ms, thus we double (400ms) it to compensate for the low intensity. 
    \item \textit{Position of vibrotactile units.} Many studies have been performed to determine the best place to position vibrotactile feedback units depending on the desired objective and the body segments. Special attention is paid to avoid giving vibrations in bony areas, placing the units aside from the spinal cord and joints, as vibrations on the bones are felt throughout the body segment and perceived as uncomfortable  \cite{Zeagler2017,Floor2018}. Moreover, according to \cite{Shull2015}, feedback units should be positioned near the body joints to be guided. Social acceptability is also a key point when considering feedback unit placement \cite{Zeagler2017,Floor2018}, trying to avoid the less accepted areas. 
    Following this criteria, we positioned the ErgoTac devices to the selected joints for the targeted task action (see Experimental protocols section).  
    \label{sec:positionofunit}
\end{itemize}

\begin{figure}
    \centering
    \includegraphics[width = 0.9\columnwidth]{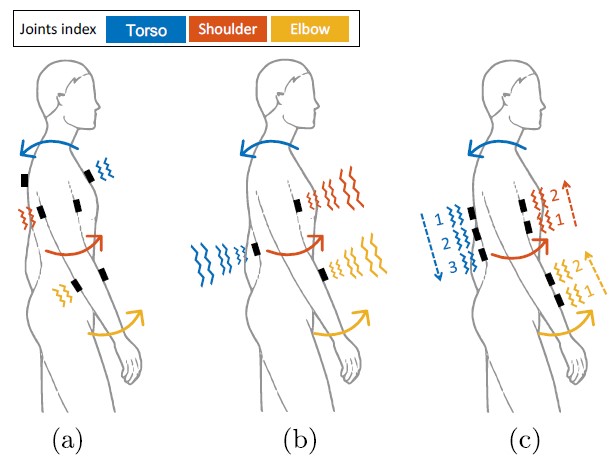}
    \caption{Illustration of the three feedback modalities: (a) SPOT, (b) RAMP, and (c) PATTERN. The black boxes represent the ErgoTac units. The triangular wave lines symbolise vibrations, where larger lines correspond to higher vibration levels. The solid arrows represent the desired movement of the torso and arm segments, respectively. In (c), the numbers and dashed arrows define the vibration sequence (Forward: 1$\longrightarrow$3 at Torso, 1$\longrightarrow$2 at Shoulder and Elbow, respectively, Backward: 3$\longrightarrow$1 in Torso, 2$\longrightarrow$1 in Shoulder and Elbow, respectively).}
    \label{fig:modality}
\end{figure}

Taking into account these factors, three feedback modalities are designed, compared, and evaluated in this work: i) SPOT; ii) RAMP; iii) PATTERN. An illustration of the ErgoTac placement and vibration styles for each modality is presented in Fig.~\ref{fig:modality}. The specifications and differences between these modalities are described below.
\subsubsection{SPOT}
This modality uses two ErgoTac units per joint: i) two on the opposite sides of the torso (chest and upper back) at the T2 level; ii) two on the upper arm at front and back sides for the shoulder; iii) two on the forearm also at front and back sides for the elbow. As depicted in Fig.~\ref{fig:modality}~(a), the desired direction is given as a repulsive vibration feedback (i.e., when feeling a vibration in one ErgoTac unit, the subject has to move the respective joint in the opposite direction) \cite{Floor2018}. 

\subsubsection{RAMP}
This modality only needs one ErgoTac unit per joint. In this case, the desired direction is given by an increasing or decreasing vibration level in each ErgoTac unit, as shown in Fig.~\ref{fig:modality}~(b). Thus, if the vibration level increases, the subject has to move the respective joint in the direction of the increment. Conversely, if the vibration level decreases, the subject has to move the respective joint in the opposite direction.

\subsubsection{PATTERN}
This modality uses three ErgoTac units for the torso and four for the arm (two units per joint). 
The number of ErgoTac units is chosen based on literature studies for the lower back \cite{Floor2018} and as a trade-off between user comfort and density of feedback information \cite{Shull2015}.
The subject has to move the respective joint following the direction given by the pattern, as shown in Fig.~\ref{fig:modality}~(c). In this figure, the numbers and dashed arrows represent the direction of the vibration sequence (pattern). For example, in the case of the torso, the ErgoTac units vibrate sequentially from number 1 to number 3, which indicates that the subject has to move the torso backward, as indicated by the solid arrow. 
Distance between the vibrotactile units is set to 5 cm based on previous studies. For instance, in \cite{Filosa2019} it was found that the best distance for comfort and to be able to distinguish the vibrations in an abdomen belt is of 5.5 cm. Similarly, in \cite{Aggravi2016} they state that the minimum distance in the forearm between two stimuli to be differentiated is about 35 mm and in \cite{Floor2018} 5 cm was found to be the optimal spacing between feedback units in the lower back. Smaller distances may be perceived as a spot vibration with a large surface \cite{Floor2018}. 

Algorithm~\ref{alg:feedback} presents the pseudocode of the aforementioned feedback modalities. This algorithm is simplified for the sake of clarity. Here, the feedback modality $f$ is chosen at the beginning of the task. Then, a while loop is executed until the user reaches the desired posture. In practice, a dead-band of 5\% is set to overcome small joint angle variations and sensors noise. 
At each step of the loop, the algorithm computes the error magnitude $\epsilon$ between the current configuration $\boldsymbol{q}_c \in \mathbb{R}^3$ and desired configuration $\boldsymbol{q}_d \in \mathbb{R}^3$ for the each $j$-th joint, respectively, ($j = [1, 2, 3] \in \mathbb{R}^3$, the three joints considered in this work: torso, shoulder, and elbow) as the absolute error normalised by the maximum error ${\xi}_j$, namely
\begin{equation}
    {\epsilon}_{j} = \frac{|{q}_{c|j}-{q}_{d|j}|}{{\xi}_{j}},
    \label{eq:error}
\end{equation}
where ${q}_{c|j}$ and ${q}_{d|j}$ are the current and desired configuration, respectively. The maximum error ${\xi}_{j}$ is chosen in accordance with the motion range of each joint as follows: the maximum error of the back, shoulder and elbow is $90^\circ$, $180^\circ$, and $145^\circ$, respectively. 
Next, the joint $j$ is selected as the one with the maximum error magnitude. Consequently, the vibration level $l$ is computed according to ${\epsilon}_{j}$ (i.e., the higher the error magnitude, the higher the vibration level). Then, the vibrotactile feedback is sent to the ErgoTacs based on the feedback modality and desired direction as described above. 

\begin{algorithm}
\caption{Directional Vibrotactile Feedback}
\label{alg:feedback}
\begin{algorithmic}
\State $f \leftarrow$ select\_feedback\_modality;
\State ${\epsilon}_{j} \leftarrow$ compute\_error;
\While {${\epsilon}_{j}$ is bigger than $5\%$}
    \State $l \leftarrow$ select\_level;
    \If{$f$ == SPOT}
        \If{${q}_{c|j} > {q}_{d|j}$} 
            \State direction = \textit{forward};
        \Else   
            \State direction = \textit{backward};
        \EndIf
        \State send\_vibration($j$, direction, $l$);
    \ElsIf{$f$ == RAMP}
        \If{${q}_{c|j} > {q}_{d|j}$} 
            \State send\_increasing\_vibration($j$, $l$);
        \Else   
            \State send\_decreasing\_vibration($j$, $l$);
        \EndIf
    \ElsIf{$f$ == PATTERN}
        \If{${q}_{c|j} > {q}_{d|j}$} 
            \State direction = \textit{forward};
        \Else   
            \State direction = \textit{backward};
        \EndIf 
        \State send\_pattern\_vibration(direction, $j$, $l$);
    \EndIf
\EndWhile
\end{algorithmic}
\end{algorithm}

\section{Ergonomics framework}
\label{sec:ergonomics_framework}

Once the feedback modalities will be tested and evaluated (Section \ref{Experiments}), the most effective solution will be validated within an ergonomic optimisation framework.

In \cite{kim2018ergotac} the design of the ErgoTac device was presented with the aim to improve human postures while performing a heavy material handling task.
The desired postures were determined based on the overloading joint torques method originally proposed in \cite{kim2017real}. This method accounted online for the torque variations induced on the human main joints by an external heavy load. The vibrotactile amplitudes were tuned depending on the level of the overloading effect in the considered joints, providing a higher amplitude for higher torques and vice-versa. Successively, the subjects were asked to  adjust their postures relying on their own intuition to minimise those vibrations (and achieve postures with less overloading torques).  
Unlike the previous approach, in this paper, human intuition is replaced by accurate feedback modalities, following a similar optimisation proposed in \cite{kim2017anticipatory}. A brief explanation of this optimisation process will be provided as follows.

The first step is the calculation of the overloading joint torque that is based on the displacement of the centre of pressure (CoP), computed from the difference between an estimated one and a measured one. The estimated CoP vector can be obtained by taking advantage of the statically equivalent serial chain (SESC) technique presented in \cite{cotton2009estimation}. On the other hand, the measured CoP vector can be collected using an external sensor system. If no interactions of the human with the external environment (or with a tool/object) occur, the estimated CoP vector is comparable to the measured one. Conversely, whether an external load is applied on the human body, the two vectors differ and the overloading joint torque vector can be estimated accordingly. Details of the method can be found in \cite{kim2017real}.

At this stage, to reduce the risk of injuries in human joints due to the overloading effect, the body configuration that minimise such torques can be obtained. To this end, an optimisation problem can be designed by setting as the objective function the sum of the weighted norms of the overloading joint torque, which depends on the human joint angles vector, subject to nonlinear inequality constraints:
\begin{align}
	\underset { \boldsymbol{q}_h }{ \text{ min } } f(\boldsymbol{q}_h) & = \frac{1}{2}\sum_{k=1}^{n_{j}}\omega_{k}|{{\tau}}_{ k }(\boldsymbol{q}_h)|^{2},\label{eq:optimisation}\\ \text{subject to: } & 
	\boldsymbol{q}_{\text{min}} \leq \boldsymbol{q}_h \leq \boldsymbol{q}_{\text{max}} , \label{eq:const_bound}\\ &
	\boldsymbol{h}_{\text{stable}} (\boldsymbol{q}_h) \leq 0,\label{eq:const_cop}\\&
	\boldsymbol{h}_{\text{task}} (\boldsymbol{q}_h) \leq 0  \label{eq:const_task},
\end{align}
where $\boldsymbol{q}_h$ is the current human joint angle when the optimisation started, $n_{j}$ is the number of joints, ${\tau} _{ k }(\boldsymbol{q}_{h})$ is the $k$-th joint overloading torque, $\omega_{k}$ is a weight associated with the joint $k$, and $\boldsymbol{h}$ are inequality box constraints. 
The weights $\omega_{k} \, > 0$ are introduced to set priorities among the joints, namely to pay more attention within the quadratic optimisation process to those ones that may be more prone to risks for a specific task. 
On the other hand, the constraints that are considered in the proposed optimisation procedure will be illustrated hereafter.

To ensure that the body configuration resulting from the optimisation is feasible and safe, \eqref{eq:const_bound} expresses a boundary condition on the joint angles, which are restricted within the human body joint physiological limits, represented by lower ($\boldsymbol{q}_\text{min}$) and upper ($\boldsymbol{q}_\text{max}$) boundaries, whose values can be found in literature \cite{whitmore2012nasa}.
The second constraint is associated with postural stability. A set of inequality constraints is considered in \eqref{eq:const_cop} to ensure that the position of CoP exists only within the convex hull of the contact points (i.e. within the support polygon of feet). Accordingly, the inequality constraint \eqref{eq:const_cop} can be formulated as
\begin{align}
\boldsymbol{h}_{\text{stable}} (\boldsymbol{q}_h) := 
\hat{\boldsymbol{C}}_{P_{wo}}(\boldsymbol{q}_h) - \text{conv}{\{\boldsymbol{p}^{x,y}_{F|j}}\} \leq 0,
\end{align}
\noindent where $\hat{\boldsymbol{C}}_{P_{wo}}(\boldsymbol{q}_h)$ is the CoP model obtained with the SESC technique, $\text{conv}{\{\boldsymbol{p}^{x,y}_{F|j}}\}$ is the convex hull including each possible $j$-th contact point $\boldsymbol{p}^{x,y}_{F}$ and it can be computed through the forward kinematics of the feet.  
Finally, \eqref{eq:const_task} expresses an inequality constraint related to the task requirement. During the experiments, subjects were required to hold a heavy object and change the body configuration, while keeping a similar height as the initial height of the object. Accordingly, \eqref{eq:const_task} can be defined as
\begin{align}
\boldsymbol{h}_{\text{task}} (\boldsymbol{q}_h) := | \boldsymbol{z}_{\text{obj}}(\boldsymbol{q}_h) | - \boldsymbol{z}_{\text{th}} \leq 0,
\end{align} 
\noindent where $\boldsymbol{z}_{\text{obj}}(\boldsymbol{q}_h)$ is the z-coordinate of the object's position (i.e., the height) that can be computed using forward kinematics while $\boldsymbol{z}_{\text{th}}$ is the z-coordinate position threshold.   

Once the human optimal body configuration is computed using \eqref{eq:optimisation}, $\boldsymbol{q}_{d} = [q_1 \cdots q_j]$ is defined (see Algorithm~\ref{alg:feedback}) and ErgoTac can be employed to assist the human subjects to achieve a more ergonomic condition.

\section{Experimental evaluation}
\label{Experiments}
The whole experimental analysis was carried out at Human-Robot Interfaces and Physical Interaction (HRI$^{2}$) Lab, Istituto Italiano di Tecnologia, Genoa, Italy, in accordance with the Declaration of Helsinki, and the protocol was approved by the ethics committee Azienda Sanitaria Locale (ASL) Genovese N.3 (Protocol IIT\_HRII\_ERGOLEAN 156/2020). Participants were students and researchers with no or limited experience of industrial work. Written informed consent was obtained after explaining the experimental procedure and a numerical ID was assigned to anonymise the data. The experimental evaluation was performed in two different steps. First, the performance of the three directional feedback modalities described earlier (i.e., SPOT, RAMP, and PATTERN) was tested when guiding the users towards the desired configuration. Next, the feedback modality with the best performance was employed in a more advanced second protocol in which an ergonomic framework provided the optimal configuration to minimise the overloading joint torques. 

\subsection{Experimental protocols}
\label{sec:exp-protocols}
Two different experimental protocols were implemented for the first and second set of experiments and are described hereafter. 

\subsubsection{Directional vibrotactile feedback modalities}
In this first protocol, both single and multi-joint feedback cases were considered separately to evaluate and compare the different modalities. For the single-joint experiment, the torso was selected as the target segment. Hence, only the lower back joint was contemplated, being one of the most affected joints by MSDs arising from industrial activities \cite{hossain2018prevalence}. This single-joint test aimed to give an idea of the feedback modalities' performance with the least possible disturbance. Instead, the multi-joint experiment aimed to compare the intuitiveness of the feedback when several joints were involved. Thus, this experiment allowed the evaluation of the three modalities in those tasks where the feedback can be less clear due to the multiple targets. For this purpose, the shoulder and elbow joints were chosen since both correspond to the arm segment and are commonly involved in manipulation tasks (see also the last paragraph, third bullet \ref{sec:positionofunit}, ``Position of vibrotactile units''). 

\begin{figure}
    \centering
    \includegraphics[width = 0.95\columnwidth]{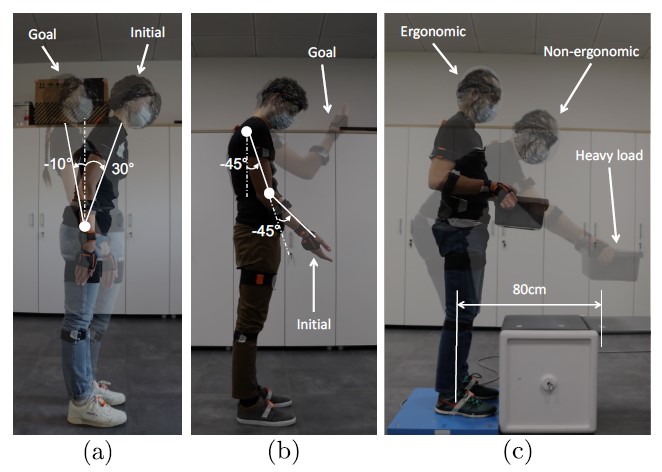}
    \caption{Sequences of pictures taken during the experiments: (a) and (b) feedback modality tests for the torso and arm, respectively; and (c) ergonomic postural adjustment with the external load at a distance of 80 cm.}
    \label{fig:experiments}
\end{figure}

Fig.~\ref{fig:experiments}~(a) and~(b) show two subjects during the experiments for the torso and arm, respectively. The subject had to move according to the vibrotactile feedback guidance towards the assigned configuration. Once the desired configuration was reached (allowing 5\% threshold error for every joint), no vibration was conveyed. A trial was considered complete when the subject announced it to the experimenter. Subsequently, the subject had to stop and wait in the current pose for the next desired pose and the new vibrotactile directional feedback. This process was applied for three consecutive configurations. In the single-joint experiment, the three desired angles for the torso were: $-10^\circ, 30^\circ, 60^\circ$. Similarly, in the multi-joint experiment, the three configurations for the arm were defined with the following desired angles for the shoulder: $10^\circ, -45^\circ, -90^\circ$, and the elbow: $-45^\circ, -90^\circ, -125^\circ$. Subjects were asked to carry out both experimental sets (torso and arm) consecutively, having few minutes to recover between them. Besides, for proper statistical evaluation, the order among the experimental sets (torso, arm) and the targeted postures were randomised among subjects. A total of fifteen subjects (age: $27.1 \pm 2.2$ years; mass: $67.8 \pm 14.5$ Kg; height: $175.9 \pm 10.4$ cm)\footnote{Subject data is reported as: mean  $\pm$ standard deviation.} participated to these experiments. In this group of subjects, 60\% were male, and 40\% were female.

\subsubsection{Ergonomic postural adjustment}
Regarding the second protocol, a more complete multi-joint case was analysed and the best performing feedback modality from previous experiments was utilised. Hence, the three joints (torso, shoulder, and elbow) were considered simultaneously. 
The aim of this experiment was to test the feedback modality in a potential ergonomic integrated use-case. To do so, the selected feedback modality was integrated within an ergonomic framework (described in section \ref{sec:ergonomics_framework}) to guide a user performing a task in a non-ergonomic posture towards an optimal ergonomic one. 

A heavy lifting task was chosen as it represents one of the most common tasks causing injuries in industrial scenarios \cite{brown1973lifting} and it involves all of the mentioned joints. Fig.~\ref{fig:experiments}~(c) shows one particular subject performing the second protocol task. The subject held a heavy object (4 kg) in a non-ergonomic posture and was guided via ErgoTac to the optimal configuration to minimise the overloading joints torques. Three different initial configurations (i.e., three different distances from the object placement) were considered. The subject started the experiment by holding the object at distances of 0.2 m (condition 1), 0.5 m (condition 2), and 0.8 m (condition 3), respectively, w.r.t. the global frame (i.e. the right heel position), as depicted in Fig.~\ref{fig:experiments}~(c). In addition to the indices collected at the previous protocol (described in section~\ref{subsubsec:indexes}), a statistical analysis was employed to find if there was a significant reduction of the overloading joint torques among the initial and final configurations.
This protocol considered a total of five subjects (age: $27.6 \pm 1.5$ years; mass: $69.4 \pm 15.7$ Kg; height: $175.6 \pm 8.4$ cm). In this group of subjects, 60\% were male, and 40\% were female.

\subsection{Experimental setup}
To measure the human kinodynamics (e.g., joint angle, Ground Reaction Force (GRF), CoP, etc.), the subjects wore the MVN Biomech suit (Xsens Technologies BV) and stood on a Kistler force plate (Kistler Holding AG). 
The ErgoTac devices were placed on the participants' skin at the considered segments (i.e., arm and torso).
The online movements and ergonomics poses (i.e., the overloading joint torques and its optimisation) were calculated using the collected sensory data. Therefore, the vibrotactile guidance (see section \ref{subsec:feedback_methods}) was defined through comparison with the subject's current configuration and the desired configuration.
The feedback module calculated the needed vibrotactile amplitudes and modalities, which were defined as in Algorithm \ref{alg:feedback} and then sent to ErgoTac.
ErgoTac devices were connected to the RF-module via Bluetooth protocol with its own specified address. The sensory data was executed at 1 kHz frequency and communicated with the feedback module at 10 Hz via the Robot Operating System (ROS).

\subsection{Measurements and indices}
\label{subsubsec:indexes}
The following indices were used to evaluate the performance of the directional vibrotactile feedback modalities in terms of physical and cognitive aspects:
\begin{itemize}

    \item For all cases: 
        \begin{itemize}
            \item \textbf{Confusion index $\mathcal{C} [\%]$}: Percentage of time in which the subject correctly followed the guidance vs the time they followed the opposite direction to the desired one. This index was normalised w.r.t. time and distance. Hence, it only rates moving direction w.r.t. the desired position.
            \item \textbf{Success ratio $\mathcal{S} [\%]$}: The number of cases where the desired position was reached vs not reached. To verify if the desired position was reached, each point of the last 2 seconds of execution was checked for those that satisfied the condition: ${\epsilon}_{j}<5\%$ (\ref{eq:error}).  
        \end{itemize} 
        
    \item For those cases in which the desired position was reached:
        \begin{itemize}
            \item \textbf{Reaching time $\Delta t [sec.]$}: Refers the duration of the reaching movement to a desired position: $\Delta t = {t_f-t_i}$, where $t_i$ is the initial time and $t_f$ was the first point in time where ${\epsilon}_{j}<5\%$ was satisfied within the last 2 seconds of execution.
            
            \item \textbf{Angular distance $\boldsymbol{\Theta} [deg]$}: Refers to the total travelled angular distance for the joint throughout the reaching movement to the desired position during time $\Delta t$.
            
            \item \textbf{Reaching velocity $\boldsymbol{v} [deg/s]$}: The average velocity the user employed to go from the initial configuration $\boldsymbol{q}_i$ to the final configuration $\boldsymbol{q}_f$ computed as
            \begin{equation}
            \boldsymbol{v} = \frac{|\boldsymbol{q}_f-\boldsymbol{q}_i|}{\Delta t},
            \end{equation}
            where  $\boldsymbol{q}_f$ is the configuration that first that satisfied the following condition: all joints' errors ${\epsilon}_{j}$ are less than 5\% in the last 2 seconds of execution.
        \end{itemize}
        
    \item For those cases in which the desired position was NOT reached:
        \begin{itemize}
            \item \textbf{Final error} $\boldsymbol{\epsilon} [\%]$: Refers to the final error that considered to the minimum $\epsilon_{j}$ for the last 2 seconds recorded. 
        \end{itemize}
        
    \item For all experiments  (focusing on usability):
         \begin{itemize}
            \item \textbf{Single Easy Question (SEQ)}: A post-task single-question  (``Overall, how difficult was the tasks with \dots ?'')  measuring users' perception of usability based on the last attempted task. The score was rated from 1 (``very hard'') to 7 (``very easy'') \cite{sauro2009comparison}. 
            \item \textbf{System Usability Scale (SUS)}: ten different questions that addressed the usability and learn ability of a system \cite{brook1996quick}.
        \end{itemize}  
        
    \item Additional index for the ergonomics test:
         \begin{itemize}
            \item \textbf{Decrement ratio} $\mathcal{D} [\%]$: Refers the reduction rate of the overloading joint torque in each joint, computed as
            \begin{equation}
                \mathcal{D}_{i} = \frac{{{\tau}}_{ i|f } - {{\tau}}_{ i|i }}{{{\tau}}_{ i|f }},
            \end{equation}
            where ${\tau} _{ i|f }$ and ${\tau} _{ i|i }$ is the $i$-th joint overloading torque in the final and initial phase, respectively. It is important to note that desired final body conﬁgurations when using the postural optimisation were not the same for all subjects.
        \end{itemize}   
\end{itemize}
A statistical method was employed to compare the significant differences of the selected indices between the feedback modalities. Statistical differences were tested with analysis of variance (ANOVA). In this analysis, the level of statistical significance used was $0.05$. All data processing was performed in MATLAB software.

\begin{figure*}
    \centering
    \includegraphics[width = 0.9\textwidth]{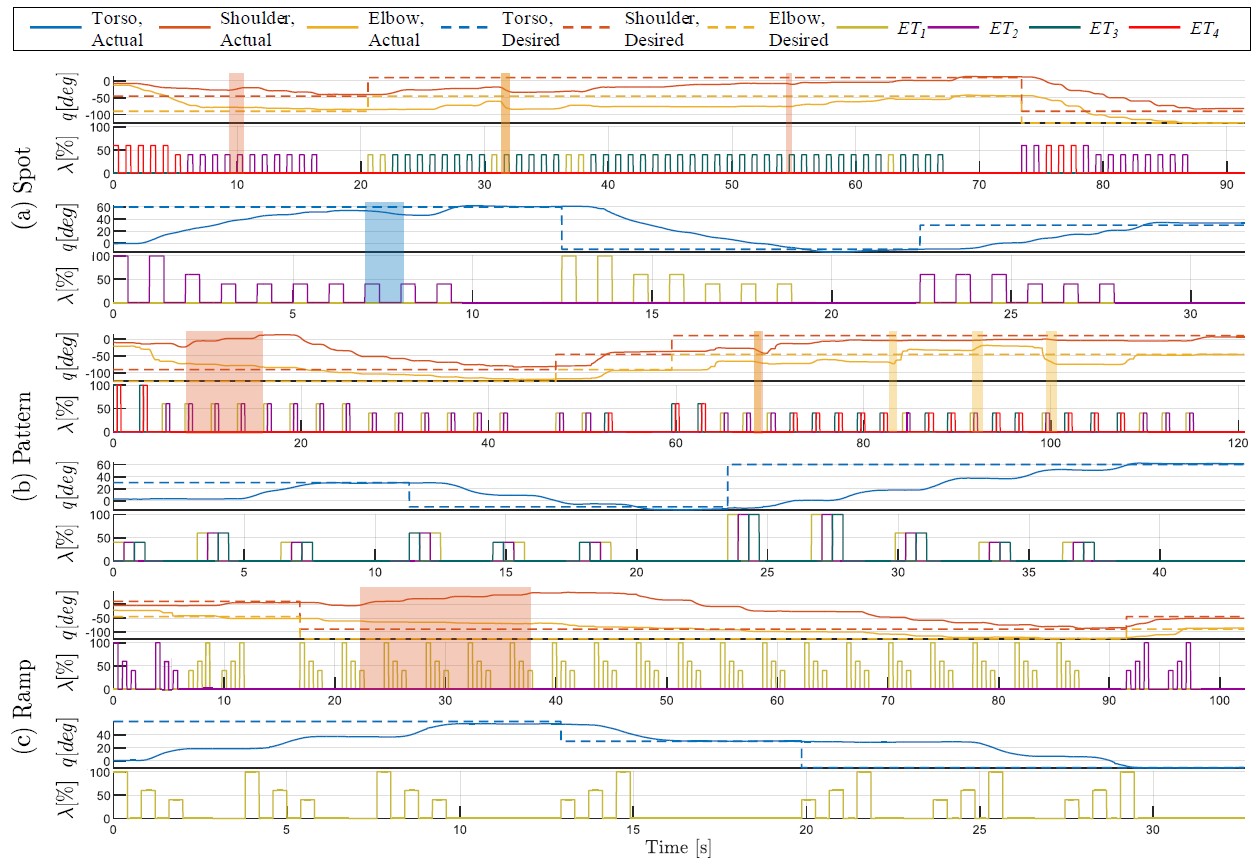}
    \caption{Excerpt of data from one particular subject collected during the modality experiment. For the three feedback modalities (a) Spot, (b) Pattern, and (c) Ramp, each subplot represents the variation of the actual (solid line) and desired angles (dashed line) of the arm (top) and torso (bottom) joints. 
    The subplot at the bottom of each row provides the amplitude of vibration ($\lambda$) of each ErgoTac device. Note that only four ErgoTac devices ($ET_{1, \ldots, 4}$) were used simultaneously in this set of experiments, as the arm and torso tests were carried out separately.
    The semi-transparent red, yellow and blue areas determine when the subject misunderstood the feedback and moved in a wrong direction the shoulder, elbow or torso, respectively.}
    \label{fig:example}
\end{figure*}

\subsection{Feedback modality test results}

Fig~\ref{fig:example} illustrates the experimental results of the feedback modality test of one particular subject while following the vibrotactile feedback guidance in real-time. This figure exposes the subject's reaction in terms of joint configuration variation during motion. 
For each feedback modality: (a) SPOT, (b) PATTERN, and (c) RAMP; the results of the arm and torso experiments are represented in the two top and bottom rows, respectively. 

At the beginning of the experimental stage, the initial configurations corresponded to considerable differences between actual and desired angles. Therefore, Algorithm \ref{alg:feedback} commanded a high amplitude to the corresponding device, providing clear vibration feedback. The vibrotactile feedback algorithm kept sending vibration stimulus until when the current configuration was the same with the desired configuration (i.e., $\boldsymbol{q}_c = \boldsymbol{q}_d$ with considering the dead-band of 5\% error). It should be noted that, once the desired configuration was reached and the vibration stopped, the subject still needed some time to become fully aware of being in the correct position. 
For instance, in the SPOT test in Fig.~\ref{fig:example}~(a), the vibration stimulus on the arm segment stopped at $18$ s; however, the subject perceived it at $20$ s.

Table \ref{tab:modalities_results} presents the overall experimental results of the feedback modality test, considering the performance indices mentioned in section~\ref{subsubsec:indexes} for all subjects ($n=15$). These indices were averaged across subjects and the statistical significance was computed to the null hypothesis of similarity of feedback modalities. 
Separate repeated-measures ANOVAs with Bonferroni corrections for multiple comparisons ($\alpha=0.05$) were applied to determine the effect of feedback modalities on each dependent variable: confusion index, success ratio, reaching time, angular distance, final error, SEQ and SUS score, respectively. Pairwise comparisons were conducted by using post-hoc paired t-tests.

The overall final errors are less than 9\%, indicating that the subjects can reach the desired position by the guidance of all the feedback modalities. Still, the fact that some are above the established 5\% threshold indicates that some subjects perceived to be in the desired position earlier than it actually happened. The overall response on the torso (2.27\%) had lower angle errors than the one on the arm (Shoulder: 5.58\% and elbow: 5.18\%). The mean final error for the overall joints reports a lower value in SPOT (4.09\%) than the other feedback modalities (RAMP: 6.89\% and PATTERN: 5.18\%). Statistical results show no significant differences (Torso--$p=0.07$, Shoulder--$p=0.61$, and Elbow--$p=0.13$), indicating that, in terms of position error, the proposed feedback modalities had similar performance.

\begin{figure}
\begin{subfigure}[b]{1\columnwidth}
    \centering
    \includegraphics[width=0.9\linewidth]{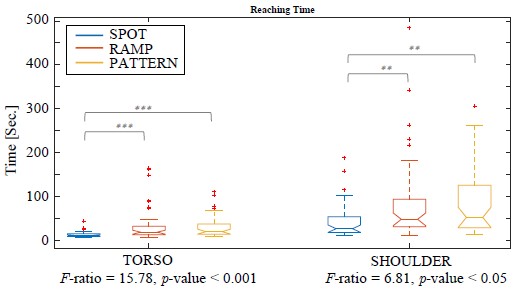}
    \caption{}
    \label{fig:result_confusion_index}
    \vspace{-0.0cm}
\end{subfigure}
\begin{subfigure}[b]{1\columnwidth}
    \centering
    \includegraphics[width=0.9\linewidth]{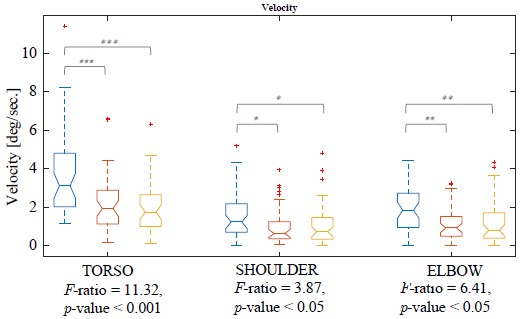}
    \caption{}
    \label{fig:result_velocity}
    \vspace{-0.0cm}
\end{subfigure}
\begin{subfigure}[b]{1\columnwidth}
    \centering
    \includegraphics[width=0.9\linewidth]{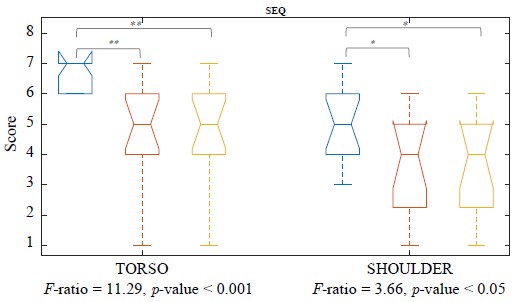}
    \caption{}
    \label{fig:result_sus_mod}
    \vspace{-0.0cm}
\end{subfigure}
    \caption{Statistical results of the reaching time, velocity and SEQ for fifteen subjects during the experiment with ErgoTac between the three modality conditions (SPOT, RAMP and PATTERN). Asterisks indicate the level of statistical signiﬁcance after post-hoc tests: $^{\ast} p <0.05$, $^{\ast\ast} p <0.01$ and $^{\ast\ast\ast} p <0.001$.}
    \label{fig:results_modalities}
    \vspace{-0.5cm}
\end{figure}

Interestingly, when performing ANOVA on different modalities significant differences were found in reaching time $\Delta t$, velocity $\boldsymbol{v}$ and SEQ of all joints ($p<0.05$), respectively. In such indices, the post hoc t-test revealed that the SPOT feedback modality significantly increased the effectiveness in several indices compared to the other modalities RAMP and PATTERN, respectively ($p<0.05$), as reported in Table \ref{tab:modalities_results} and Fig.\ref{fig:results_modalities}. 
A bar graph of the average of the reaching time, velocity and SEQ for all subjects can be seen in Fig.\ref{fig:results_modalities}. For all the three indices, and for all the considered segments, there were statistically significant differences in the indices between SPOT and RAMP, or PATTERN feedback modality, respectively. 
On the other hand, the angular distance index reported notable differences in torso and shoulder ($p<0.05$), while for the elbow, the difference was not remarkable ($p=0.09$). Besides, the post hoc t-test on the angular distance index showed that the movements when using SPOT were lower than when using the other modalities ($p\leq0.05$). Note that the comparison results of \textrm{SPOT} and \textrm{RAMP} in the upper arm and forearm were $p=0.05$. In fact, RAMP and PATTERN modalities increased the overall travelled distances and time to reach the desired position: $\textrm{SPOT} = 28.35$ s, $\textrm{RAMP} = 60.24$ s. and  $\textrm{PATTERN} = 54.83$ s. 
The confusion index depicted remarkable differences among the feedback modalities for the torso experiment ($p<0.001$). This difference was also notable in the pairwise comparison for the SPOT modality ($p<0.001$). However, no important effect was revealed on the arm (shoulder--$p=0.29$ and elbow--$p=0.12$).

The average SUS scores were 77.33 (SPOT), 56.83 (RAMP), and 56.83 (PATTERN), respectively. The SUS score of SPOT feedback modality indicated that user acceptance is ``Excellent'' (adjective rating) or ``Acceptable'' (acceptability rating), while user acceptance of the other modalities according to the SUS scores reported is ``OK'' or ``Marginal''. Moreover, SPOT showed a significant main effect in the pairwise comparison (SPOT versus RAMP $p<0.05$ and SPOT versus PATTERN $p<0.05$). 
These results implied that SPOT modality was more convenient than the two other feedback modalities and enhanced the vibrotactile feedback framework's performance. Hence, the SPOT modality was selected as the best feedback modality and exploited in the ergonomic postural evaluation.       

\begin{table*}[]
\caption{\label{tab:modalities_results}Experimental results of fifteen subjects. Differences between the three feedback modalities (SPOT, RAMP and PATTERN) and the results from the repeated-measured ANOVAs.}
\begin{tabularx}{\textwidth}{@{}Llcccccccc@{}}
\hline
\multicolumn{1}{l}{}                                &          & \multicolumn{3}{c}{Mean (standard deviation) results}           & \multicolumn{2}{c}{ANOVA results} & \multicolumn{3}{c}{Pairwise Comparisons}                \\ \cline{3-10} 
\multicolumn{1}{l}{}                                &          & {SPOT (S)}& {RAMP (R)}         & {PATTERN (P)}   & $F$-ratio  & $p$-Value            & S vs. R               & S vs. P               & R vs. P \\ \hline
\multirow{3}{*}{\shortstack[l]{Final error,\\ $\boldsymbol{\epsilon}$ {[}\%{]}}}  & Torso     & $1.96\pm1.26$& $2.27 \pm1.34$& $2.64    \pm1.49$& 2.73       & 0.07                 & 0.26                  & 0.02                  & 0.23    \\
                                                    & Upper arm & $5.14 \pm6.91$& $6.89 \pm11.39$& $5.58    \pm6.54$& 0.50       & 0.61                 & 0.39                  & 0.76                  & 0.51    \\
                                                    & Forearm    & $4.09 \pm4.54$& $8.93 \pm15.72$& $5.18    \pm6.31$& 2.10       & 0.13                 & 0.06                  & 0.36                  & 0.15    \\\\
\multirow{2}{*}{\shortstack[l]{Reach time,\\ $\Delta t$ {[}sec.{]}}} & Torso     & $12.70 \pm7.12$& $34.24 \pm38.98$& $30.61   \pm25.36$& 15.76      & \boldsymbol{$p < 0.001$} & \boldsymbol{$p < 0.001$}  & \boldsymbol{$p < 0.001$}  & 0.61    \\
                                                    & Arm      & $44.00 \pm39.19$& $86.24 \pm93.89$& $79.06   \pm67.88$& 6.81       & \boldsymbol{$p < 0.05$}  & \boldsymbol{$p < 0.0167$} & \boldsymbol{$p < 0.0167$} & 0.68    \\\\
\multirow{2}{*}{\shortstack[l]{Success rate,\\ $\mathcal{C}$ {[}\%{]}}}              & Torso     & $100.00 \pm.00$& $100.00 \pm.00$& $100.00  \pm.00$      & 0.37                 & -                     & -                     & -       \\
                                                    & Arm      & $88.89 \pm31.43$& $77.78 \pm41.57$& $84.44   \pm36.24$& 0.99       & \boldsymbol{$p < 0.001$} & 0.16                  & 0.54                  & 0.42    \\\\
\multirow{3}{*}{\shortstack[m]{Velocity,\\ $\boldsymbol{v}$ {[}deg/s{]}}}    & Torso     & $3.66 \pm2.08$& $2.14 \pm1.48$& $1.94    \pm1.26$& 11.32      & \boldsymbol{$p < 0.05$} & \boldsymbol{$p < 0.001$}  & \boldsymbol{$p < 0.001$}  & 0.51    \\
                                                    & Upper arm & $1.59 \pm1.15$& $1.01 \pm.93$& $1.09    \pm1.09$& 3.87       & \boldsymbol{$p < 0.05$}  & \boldsymbol{$p < 0.0167$} & 0.04                  & 0.70    \\
                                                    & Forearm    & $1.88 \pm1.18$& $1.13 \pm.92$& $1.20    \pm1.13$& 6.41       & \boldsymbol{$p < 0.05$}  & \boldsymbol{$p < 0.0167$} & \textbf{$p < 0.0167$} & 0.75    \\\\
\multirow{3}{*}{\shortstack[l]{Angular\\ distance,\\ $\boldsymbol{\Theta}$ {[}deg{]}} }      & Torso     & $60.62 \pm27.23$& $109.56 \pm103.57$& $105.00    \pm98.61$& 8.22       & \boldsymbol{$p < 0.001$} & \boldsymbol{$p < 0.0167$} & \boldsymbol{$p < 0.001$}  & 0.83    \\
                                                    & Upper arm & $88.23 \pm47.97$& $158.14 \pm224.60$& $145.77    \pm126.02$& 5.60       & \boldsymbol{$p < 0.05$}                 & 0.05                  & \boldsymbol{$p < 0.0167$}                  & 0.75    \\
                                                    & Forearm    & $141.37 \pm116.44$& $237.48    \pm302.25$&   $218.39 \pm193.89$&   2.40	&   0.09	&   0.05	&   0.03	&   0.73    \\\\ 
\multirow{3}{*}{\shortstack[l]{Confusion\\ index,\\ $\mathcal{C}$ {[}\%{]}}}          & Torso     & $31.48 \pm12.19$& $40.26 \pm8.81$& $41.63   \pm9.64$& 12.53      & \boldsymbol{$p < 0.001$} & \boldsymbol{$p < 0.001$}  & \boldsymbol{$p < 0.001$}  & 0.49    \\
                                                    & Upper arm & $38.88 \pm12.98$& $41.46 \pm13.52$& $43.09   \pm11.01$& 1.26       & 0.29                 & 0.36                  & 0.10                  & 0.54    \\
                                                    & Forearm    & $42.38 \pm11.38$& $42.71 \pm9.65$& $46.61   \pm10.72$& 2.16       & 0.12                 & 0.88                  & 0.08                  & 0.08    \\ \hline
\multirow{2}{*}{SEQ}                                & Torso     & $6.67 \pm.49$& $5.13 \pm1.60$& $5.13    \pm1.60$& 11.30      & \boldsymbol{$p < 0.001$} & \boldsymbol{$p < 0.0167$} & \boldsymbol{$p < 0.0167$} & 1.00    \\
                                                    & Arm      & $4.93 \pm1.39$& $3.60 \pm1.64$& $3.60    \pm1.64$& 3.66       & \boldsymbol{$p < 0.05$}  & 0.02                  & 0.02                  & 1.00    \\\\
SUS                                                 & -        & $77.33 \pm10.67$& $56.83 \pm24.15$& $56.83   \pm24.15$& 7.56       & \boldsymbol{$p < 0.05$}  & \boldsymbol{$p < 0.0167$} & \boldsymbol{$p < 0.0167$} & 1.00    \\ \hline
\end{tabularx}
\begin{flushleft}Bold values indicate a significant difference at $p<0.05$.
\end{flushleft}
\vspace{-0.5cm}
\end{table*}

\subsection{Ergonomics test results}
As mentioned in section \ref{sec:exp-protocols}, an experiment with five subjects was carried out to evaluate the performance of the selected feedback modality (ie. SPOT) to get the optimised ergonomic posture in the multi-joint case (i.e., torso and arm simultaneously). During these experiments, the desired angles were given by the ergonomics optimisation framework described in section \ref{sec:ergonomics_framework}. 

\begin{figure}
    \centering
    \includegraphics[width = 1.0\columnwidth]{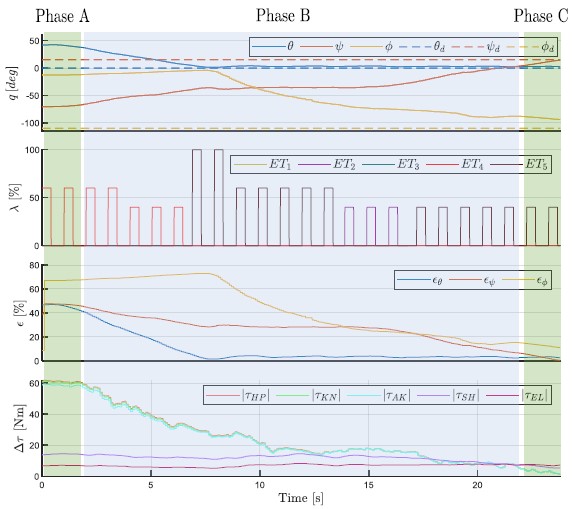}
    \caption{Excerpt of data from one particular subject collected during the ergonomic experiment in condition 2. The motion started in phase A, then the human pose is adjusted using the feedback (phase B) toward phase C. From top to bottom: Variation of actual angles, vibration levels of each ErgoTac, error magnitude, and absolute torque increments (HP:hip, KN:knee, AN:ankle, SH:shoulder, EL:elbow).}
    \label{fig:results_ergo}
    \vspace{-0.5cm}
\end{figure}

Fig.\ref{fig:results_ergo} illustrates the experiment results for one particular subject handling the heavy object at a distance of 0.5m (i.e., condition 2). The first row depicts the variations of the actual (solid line) and desired (dashed line) angles for the torso ($\theta$), shoulder ($\psi$), and elbow($\phi$). The initial conﬁguration (phase A) corresponds to a high-risk conﬁguration from the ergonomics point of view, i.e., lifting the box far from the desired (optimal) configuration, adjusting the configuration (phase B), and the final configuration (phase C) corresponds to the low-risk configuration that was adjusted through the ErgoTac feedback. The second row depicts how the vibration amplitude and direction are conveyed to the subject through the SPOT feedback modality, showing how the vibration level is reduced as the user approaches the optimal ergonomic posture. The last two plots show the error magnitude and the estimated overloading joint torques, respectively. The latter represents the absolute torque increments of the hip ($\tau_H$)\footnote{The hip corresponds to the torso in the overloading joint torques model \cite{kim2017anticipatory}}, knee ($\tau_K$), ankle ($\tau_A$), shoulder ($\tau_S$), and elbow ($\tau_E$).
ErgoTac demonstrated the advantage of providing the appropriate feedback through the SPOT modality to find an ergonomic configuration in real-time, as proved by the decrements of the overloading joint torques.

Table \ref{tab:ergonomics_results} shows the overall results for all subjects ($n=5$) in terms of the proposed indices. In addition, the decrement ratio ($\mathcal{D}$) of joint overload torque for each joint and the usability score are shown. In particular, $\mathcal{D}$ was calculated from the difference between the overloading joint torque value in the initial (phase A) and final (phase C) configuration and then averaged across all subjects for each condition.

The overall values of the indices reflected the differences among the three conditions, where performance worsened as the condition was more extreme (i.e., the worst performance was when the heavy object was at a distance of 0.8m). Generally, the retrieved indices presented higher values than in the previous experiment (i.e. the feedback modality test). However, it must be noted that the ergonomic experiment addresses more degrees of freedom (torso, shoulder and elbow), and thus, reaching the target configuration was more demanding than in the feedback modality test. 
The mean SUS score is 79.58 points, indicating that user acceptance is ``Excellent'' (adjective rating) or ``Acceptable'' (acceptability rating).

\begin{figure*}
    \centering
    \includegraphics[width=0.8\linewidth]{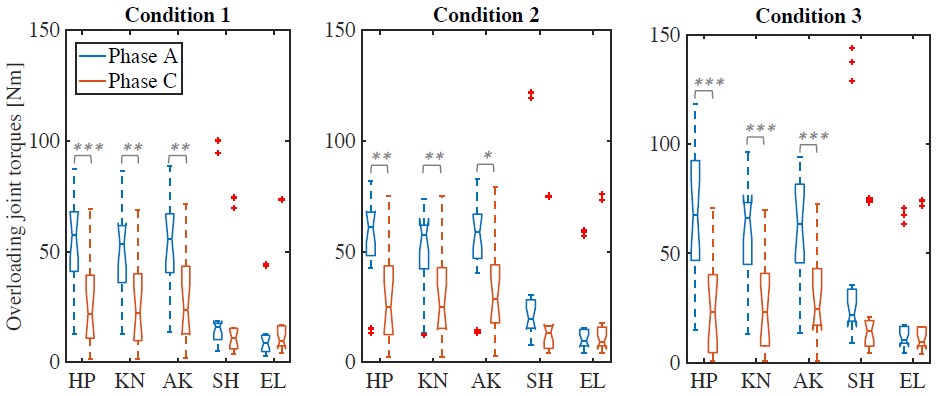}
    \caption{Statistical results of the joint overloading for five subjects during the experiments with the SPOT feedback modality by ErgoTac. The joints overloading (SH: shoulder, EL: elbow, HP: hip, KN: knee, AK: ankle) of most of the joints resulted in phase (A) are signiﬁcantly greater than phase (C), except the EL and SH joints. Asterisks indicate the level of statistical signiﬁcance after post-hoc tests: $^{\ast} p <.05$, $^{\ast\ast} p <.01$ and $^{\ast\ast\ast} p <.001$.}
    \label{fig:results_overloadingtorque}
    \vspace{-0.5cm}
\end{figure*}

From these results, it can be observed that the shoulder overloading joint torque is reduced. 
Interestingly, even the torques in hip, knee and ankle joints exhibit significant reductions in response to the feedback. 
Fig. \ref{fig:results_overloadingtorque} represents the results of statistical signiﬁcance for the overloading joint torque using the data collected in phases A and C that allow us to evaluate the ergonomic performance.
In the optimised configuration (phase C), the overloading joint torques in hip, knee and ankle report significantly decreased effectiveness compared to the higher risk configuration (phase A) ($p<0.05$). However, this main effect was not observed in the behaviours of the arm joints: the reduction rate of shoulder and elbow was, respectively:  $27.06\pm9.53\%$ ($p=0.47$), and $-36.78\pm27.74\%$ ($p=0.34$) in condition 1; $40.20\pm12.77\%$ ($p=0.23$) and $-5.71\pm14.40\%$ ($p=.71$) in condition 2; and $43.74\pm12.72\%$ ($p=0.14$) and $5.27\pm9.98\%$ ($p=0.96$) in condition 3.

\begin{table}[]
\caption{\label{tab:ergonomics_results}Experimental result of five subjects. The results are reported according to three experimental conditions (i.e., distances of the object to be lifted, see section \ref{sec:exp-protocols}).}
\centering
\begin{tabularx}{\linewidth}{@{}>{\hsize=0.6\hsize}X>{\hsize=0.8\hsize}XXXX@{}}
\hline
                                                   &          & \multicolumn{3}{c}{Mean (standard deviation) results}   \\ \cline{3-5} 
                                                   &          & {Condition 1 (0.2 m) }          & {Condition 2 (0.5 m) }          & {Condition 3 (0.8 m) }         \\ \hline
\multirow{3}{*}{\shortstack[l]{Final error,\\ $\boldsymbol{\epsilon}$ {[}\%{]}}} & Torso     & $2.14\pm1.12$& $2.04 \pm1.46$& $2.62 \pm2.14$ \\
                                                   & Upper arm & $4.90 \pm2.89$& $6.84 \pm6.62$& $5.24 \pm3.85$ \\
                                                   & Forearm    & $2.16 \pm1.70$& $7.34 \pm8.09$& $8.64 \pm10.78$ \\ \\
\multicolumn{2}{l}{{Reach time, $\Delta t$ {[}sec.{]}}}         & $27.60 \pm18.71$& $28.86  \pm15.35$& $56.09 \pm73.94$ \\ \\
\multicolumn{2}{l}{{Success rate, $\mathcal{C}$ {[}\%{]}}}  & $7.78 \pm41.57$& $77.78  \pm41.57$& $83.33 \pm37.27$ \\ \\
\multirow{3}{*}{\shortstack[m]{Velocity,\\ $\boldsymbol{v}$ {[}deg/s{]}}}  & Torso     & $1.13 \pm.87$& $1.85 \pm.97$ & $1.76 \pm1.00$ \\ 
                                                   & Upper arm & $2.33 \pm1.45$& $3.00 \pm1.46$& $2.87 \pm1.80$ \\
                                                   & Forearm    & $4.46 \pm3.12$& $3.61 \pm2.21$& $2.70 \pm1.96$ \\ \\
\multirow{3}{*}{\shortstack{Speed,\\ {[}deg/s{]}}}      & Torso     & $2.69 \pm1.15$& $3.16 \pm1.27$& $3.02 \pm1.06$  \\
                                                   & Upper arm & $3.15 \pm1.27$& $3.82 \pm1.42$& $3.72 \pm1.70$  \\
                                                   & Forearm    & $5.68 \pm3.29$& $5.02 \pm2.96$& $4.15 \pm1.66$ \\ \\
\multirow{3}{*}{\shortstack[l]{Confusion\\ index,\\ $\mathcal{C}$ {[}\%{]}}}         & Torso     & $46.06 \pm4.77$& $44.34  \pm5.46$& $43.16 \pm6.51$ \\
                                                   & Upper arm & $30.16 \pm13.26$& $24.88  \pm8.94$& $28.66 \pm9.92$ \\
                                                   & Forearm    & $38.41 \pm10.12$& $36.84  \pm11.24$& $36.27 \pm10.04$ \\ \hline
\multirow{5}{*}{\shortstack[l]{ Decrement\\ ratio,\\ $\mathcal{D}$  {[}\%{]}} }        & SH       & $27.06 \pm9.53$& $40.20  \pm12.77$& $43.74 \pm12.72$ \\
                                                   & EL       & -36.78$\pm$27.74& $-5.71  \pm14.40$& $5.27 \pm9.98$ \\
                                                   & HP       & $50.20 \pm31.90$& $25.36  \pm68.47$& $50.74 \pm48.78$ \\
                                                   & KN       & $48.08 \pm31.15$& $19.29  \pm75.65$& $45.23 \pm54.60$ \\
                                                   & AK       & $44.53 \pm28.41$& $15.93  \pm71.28$& $42.45 \pm51.85$ \\ \hline 
SUS                                                & -        & \multicolumn{3}{c}{$79.58 \pm12.29$} \\ \hline
\end{tabularx}
\end{table}

\section{Discussion}

From the results reported above, some significant outcomes are discussed in this section. 

A notable finding is that, generally, the arm feedback corresponds to worse rates than torso feedback. This is not an unexpected effect since the arm movement involves multiple joints, while the torso movement considers the case of a single joint. Hence, the arm task requires more attention from the user than the torso. This also implies an increase in the number of instruction following errors, resulting in a higher confusion index.  

Regarding the feedback modality test, although all can provide appropriate directional information to the subjects, the results implicate that SPOT outperforms the two others and enhanced the vibrotactile feedback framework's performance. Overall, the SPOT feedback modality produces a $53.20\%$ and $44.01\%$ improvement over RAMP and PATTERN, respectively. Consequently, it conveys higher usability experiences to the subjects to fulfil the capability of ergonomic posture guidance. 

In addition, the ergonomic feedback guidance experiments using the SPOT feedback modality on the three joints (torso, shoulder and elbow) showed that the overloading joint torques at the elbow and shoulder did not show a noticeable change in all cases, and in some of them, it demonstrated an increment of the overloading torque at the elbow joint.
Nevertheless, the overloading torques in hip (i.e., torso), knee and ankle joints also exhibited significant reductions in response to the feedback modality. This indirect effect was caused by the re-configuring of the external load. The steady external forces were distributed into the joint forces due to the arm movements causing the smaller moment arm to the lower body and consequently lower overloading joint torques in lower extremities. 
This effect also can be explained by the task constraint of keeping the object by hands at the same height and balancing the body pose during the task. Nevertheless, in general terms, it can be ensured that the proposed framework contributed to a coherent reduction of the overloading joint torques when performing the lifting task using the SPOT feedback modality in the multi-joint and -segment cases. Therefore, the proposed framework has the potential to raise awareness and provide directional guidance to reduce some occupational risks.

\section{Conclusions}
In this work, we proposed an ergonomics feedback framework with the ability to provide directional guidance at specific body segments towards an optimal ergonomic configuration. The interface integrated the ErgoTac vibrotactile device \cite{kim2018ergotac} to provide feedback within three different modalities (i.e., SPOT, PATTERN, and RAMP). 

Two experimental evaluations were carried out. The first evaluation results demonstrated that the SPOT feedback modality was the most intuitive and preferable modality to provide directional guidance. This statement was supported by both objective and subjective statistical analysis. Thus, this modality was implemented along with the ergonomic optimisation into the overall ergonomics feedback framework. The second evaluation outcomes verified the benefits of the proposed system on the minimisation of the overloading joint torques when performing a physically demanding task and, therefore, on improving the ergonomics condition of workers. 

 One potential limitation for real-life application is that the proposed ergonomic framework includes expensive devices (e.g., XSENS, IMU-based motion capture system). However, this limitation can be overcome by using more affordable motion capture systems (e.g., vision sensor) or by integrating the IMU-based motion capture sensor into the ErgoTac module.

Future work will focus on studying the multi-joint (or -segments) in a whole-body motion when performing intense tasks in real industrial demonstrations, and on the imposed cognitive burden with a higher number of sensors. Moreover, the proposed methodology will also determine an optimal frequency of vibration for ErgoTac device, and will be compared with other haptic devices that provide directional cue, such as skin stretch, slip motion and pressure-based devices.

\ifCLASSOPTIONcaptionsoff
  \newpage
\fi

\bibliographystyle{ieeetr}
\bibliography{ms.bib}

\begin{IEEEbiography}[{\includegraphics[width=1in,height=1.25in,clip,keepaspectratio]{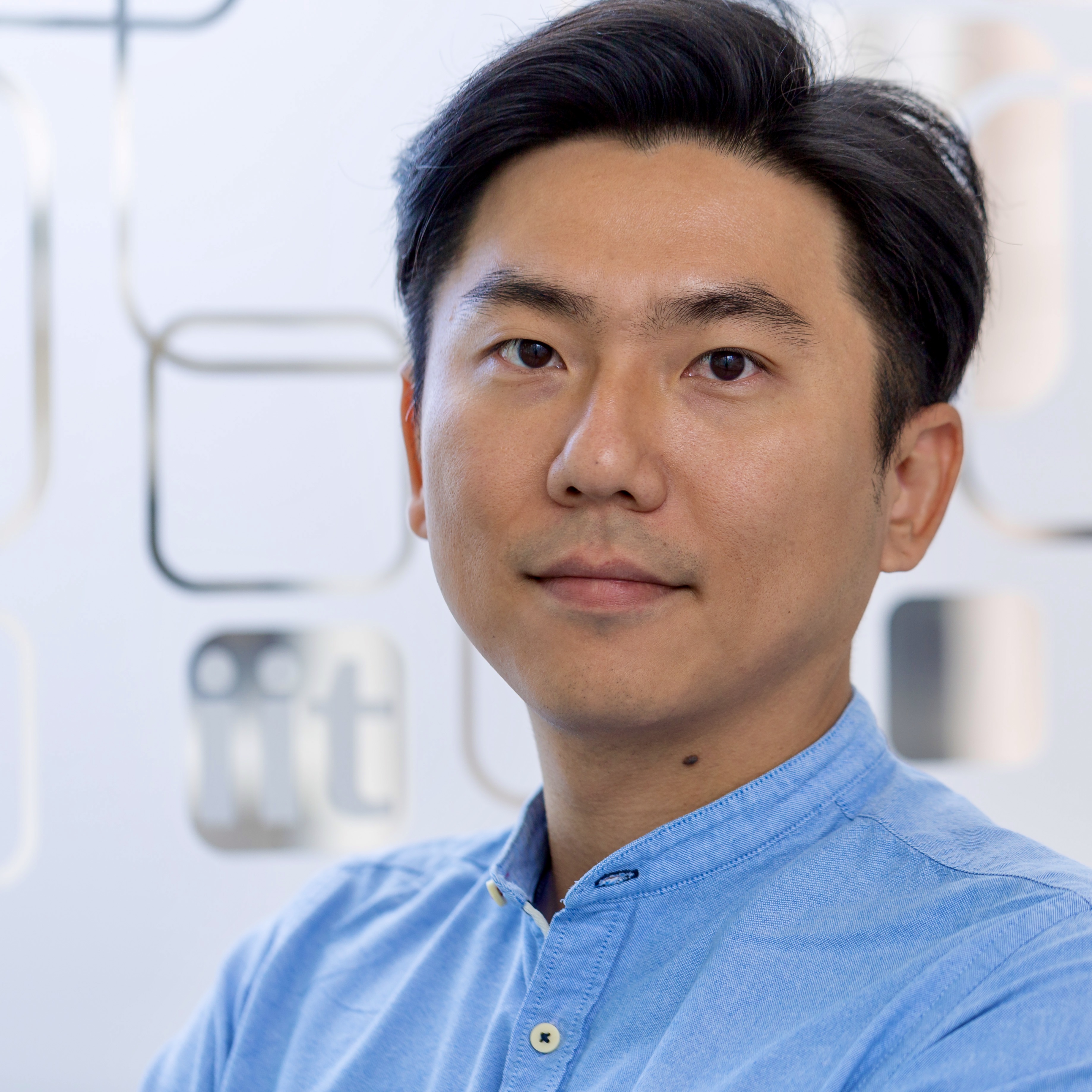}}]{Wansoo Kim}
is an assistant professor at Hanyang University ERICA, Republic of Korea. He received the B.S. degree in mechanical engineering from Hanyang University, Korea in 2008 and a Ph.D. degree in mechanical engineering from Hanyang University, Korea in 2015 (Integrated MS/PhD program). He was with Human-Robot Interfaces and Physical Interaction (HRI$^2$) Lab., Italian Institute of Technology in Genoa, Italy from 2016 to 2021. He was the winner of the Solution Award 2019 (Premio Innovazione Robotica at MECSPE2019), the winner of the KUKA Innovation Award 2018, the winner of the HYU best PhD paper award 2015, and the winner of the ICCAS best presentation award 2014. His research interests are in Physical human-robot interaction (pHRI), human-robot collaboration, Shared Control, Ergonomics, Human modelling, Feedback devices, and powered exoskeleton robot. 
\end{IEEEbiography}

\begin{IEEEbiography}[{\includegraphics[width=1in,height=1.25in,clip,keepaspectratio]{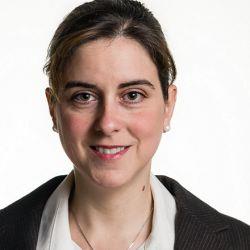}}]{Virginia Ruiz Garate} joined UWE with as a Wallscourt Fellow in Intelligent Assisted Robotics on January 2021. She was a Post- Doc at the Human-Robot Interfaces and Physical Interaction (HRI$^2$) laboratory of the Istituto Italiano di Tecnologia (IIT) in Italy. She obtained her PhD in 2016 from the Universite Catholique de Louvain (UCL) in Belgium under the EU collaborative project CYBERLEGs. 
She co-organized the RSS 2019 workshop on “Emerging Paradigms for Robotic Manipulation: from the Lab to the Productive World” from which a RAM SI developed. In the HRI2 group she worked for the SOMA project studying new grasping paradigms, and she is now researching under the SOPHIA project. Her cur- rent research interests include grasping and manipulation, bio-inspired control, assistive robotics, and human-robot collaboration.
\end{IEEEbiography}

\begin{IEEEbiography}[{\includegraphics[width=1in,height=1.25in,clip,keepaspectratio]{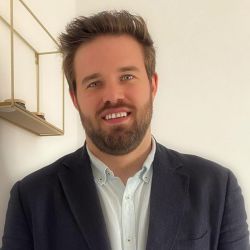}}]{Juan M. Gandarias}
is a postdoctoral researcher at the Human-Robot Interfaces and Physical Interaction (HRI$^2$) at Istituto Italiano di Tecnologia (IIT). He received the B.S., M.S., and the PhD in Mechatronics from the University of Malaga in 2015, 2017, and 2020, respectively. He is currently involved in Horizon-2020 project SOPHIA and ERC project Ergo-Lean. He has contributed to several Spanish and European projects related to search-and-rescue, physical robotic assistance and Human-Robot Collaboration in Industrial environments. 
His research interests include HRC, human modelling, and haptic perception.
\end{IEEEbiography}

\begin{IEEEbiography}[{\includegraphics[width=1in,height=1.25in,clip,keepaspectratio]{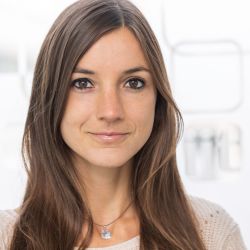}}]{Marta Lorenzini}
is a postdoctoral researcher at the Human-Robot Interfaces and Physical Interaction lab (HRI2) at Istituto Italiano di Tecnologia (IIT). She received the B.S., M.S., and Ph.D. in Department of Electronics, Information and Bioengineering from Politecnico di Milano, Milano, Italy in 2014, 2016, and 2020, respectively. She is currently involved in Horizon-2020 project SOPHIA and ERC project Ergo-Lean. She was the winner of the Solution Award 2019 (Premio Innovazione Robotica at MECSPE2019) and the winner of the KUKA Innovation Award 2018. Her research interests include human kinodynamic states real-time monitoring, human ergonomics estimation and assessment, physical human-robot interaction and feedback interfaces.
\end{IEEEbiography}

\begin{IEEEbiography}[{\includegraphics[width=1in,height=1.25in,clip,keepaspectratio]{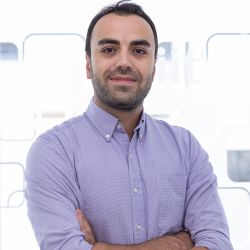}}]{Arash Ajoudani} is a tenured senior scientist at the Italian Institute of Technology (IIT), where he leads the Human-Robot Interfaces and physical Interaction (HRI²) laboratory. He received his PhD degree in Robotics and Automation from University of Pisa and IIT in 2014. He is a recipient of the European Research Council (ERC) starting grant 2019, the coordinator of the Horizon-2020 project SOPHIA, and the co-coordinator of the Horizon-2020 project CONCERT. He is a recipient of the IEEE Robotics and Automation Society (RAS) Early Career Award 2021, and winner of the Amazon Research Awards 2019, of the Solution Award 2019 (MECSPE2019), of the KUKA Innovation Award 2018, of the WeRob best poster award 2018, and of the best student paper award at ROBIO 2013. His PhD thesis was a finalist for the Georges Giralt PhD award 2015 - best European PhD thesis in robotics. He was also a finalist for the Solution Award 2020 (MECSPE2020), the best conference paper award at Humanoids 2018, for the best interactive paper award at Humanoids 2016, for the best oral presentation award at Automatica (SIDRA) 2014, and for the best manipulation paper award at ICRA 2012. 
He is currently serving as the executive manager of the IEEE-RAS Young Reviewers' Program (YRP), and as chair and representative of the IEEE-RAS Young Professionals Committee.  
His main research interests are in physical human-robot interaction, mobile manipulation, robust and adaptive control, assistive robotics, and tele-robotics.
\end{IEEEbiography}

\end{document}